\documentclass[aps,prb,twocolumn,superscriptaddress,showpacs]{revtex4-1}
\usepackage{graphicx}
\usepackage{latexsym}
\usepackage{amssymb}
\usepackage{amsmath}
\usepackage{amsfonts}
\usepackage{epstopdf}
\usepackage{bm}
\usepackage{multirow}
\usepackage{color}
\usepackage[colorlinks,linkcolor=blue,citecolor=blue,urlcolor=blue]{hyperref}
\newcommand{\ii}{\mathrm{i}}
\newcommand{\dd}{\mathrm{d}}

\renewcommand{\Re}{\mathop{\mathrm{Re}}}
\renewcommand{\Im}{\mathop{\mathrm{Im}}}
\newcommand{\vect}[1]{{\bm{#1}}}
\newcommand{\eqnref}[1]{Eq.\,\eqref{#1}}

\newcommand{\tabref}[1]{Tab.\,\ref{#1}}

\newcommand{\beq}{\begin{equation}}
\newcommand{\eeq}{\end{equation}}
\newcommand{\beqn}{\begin{eqnarray}}
\newcommand{\eeqn}{\end{eqnarray}}

\begin{document}

\title{Bona fide interaction-driven topological phase transition in correlated SPT states}
\author{Yuan-Yao He}
\author{Han-Qing Wu}
\address{Department of Physics, Renmin University of China, Beijing 100872, China}
\author{Yi-Zhuang You}
\affiliation{Department of Physics, University of California,
Santa Barbara, California 93106, USA}
\author{Cenke Xu}
\affiliation{Department of Physics, University of California,
Santa Barbara, California 93106, USA}
\author{Zi Yang Meng}
\affiliation{Beijing National Laboratory
for Condensed Matter Physics, and Institute of Physics, Chinese
Academy of Sciences, Beijing 100190, China}
\author{Zhong-Yi Lu}
\address{Department of Physics, Renmin University of China, Beijing 100872, China}

\begin{abstract}

It is expected that the interplay between non-trivial band
topology and strong electron correlation will lead to very rich
physics. Thus a controlled study of the competition between
topology and correlation is of great interest. Here, employing
large-scale quantum Monte Carlo (QMC) simulations, we provide a
concrete example of the Kane-Mele-Hubbard (KMH) model on an AA
stacking bilayer honeycomb lattice with inter-layer
antiferromagnetic interaction. Our simulation identified several
different phases: a quantum spin-Hall insulator (QSH), a
$xy$-plane antiferromagnetic Mott insulator ($xy$-AFM) and an
inter-layer dimer-singlet insulator (dimer-singlet). Most
importantly, a bona fide topological phase transition between the
QSH and the dimer-singlet insulators, purely driven by the
inter-layer antiferromagnetic interaction is found. At the
transition, the spin and charge gap of the system close while the
single-particle excitations remain gapped, which means that this
transition has no mean field analogue and it can be viewed as a
transition between bosonic SPT states. At one special point, this
transition is described by a $(2+1)d$ $O(4)$ nonlinear sigma model
(NLSM) with {\it exact} $SO(4)$ symmetry, and a topological term
at {\it exactly} $\Theta = \pi$.
Relevance of this work towards more general interacting SPT
states is discussed.
\end{abstract}

\pacs{71.10.-w, 71.10.Fd, 71.27.+a}

\date{\today} \maketitle

\section{INTRODUCTION}

The interplay between non-trivial band topology and strong
electron interaction is expected to lead to a plethora of new
physical phenomena in strongly correlated systems. Many exotic
phenomena of interacting topological insulators (TI) have been
predicted/discovered, such as topological Kondo
insulator~\cite{kaisun,kondo1,kondo2}, fractionalized
TI~\cite{balents,Maciejko2015}, interaction-reduced classification
of
TI~\cite{Fidkowski1,Fidkowski2,XLQi2013,HYao2013,SCZhang2012,ZCGu2014,
Fidkowski2013,Senthil2014,YZYou2014_TSC,YZYou2014_SPT,YZYou2014_ITI},
and interaction-driven anomalous topological order at the boundary
of TIs~\cite{TI_fidkowski1,TI_fidkowski2,TI_qi,TI_senthil,TI_max,Dual_Dirac_liquid_Wang,Duality_max}.
Besides fermionic systems, it was also proposed that bosonic
systems can also form exotic states that are similar to fermionic
TIs~\cite{Wen_SPT,Wen_SPT2}, which are generally called the
symmetry protected topological (SPT) states. Unlike their
fermionic counterparts, bosonic SPT states can only exist in
strongly interacting boson systems, and the interaction must be
carefully designed to avoid the ordinary superfluid and Mott
insulator phases. These studies have tremendously broadened our
understanding of quantum disordered states of matter and revealed
the fundamental role topology plays in condensed matter systems.

Quantum phase transition between different stable quantum
disordered phases is another important subject, and in general it
can be very different from the standard Ginzburg-Landau (GL) phase
transition paradigm. For example, one expects a phase transition between a $(2+1)d$ topological ordered state ($Z_2$ spin liquid~\cite{Isakov2006}) and an conventionally ordered phase (superfluid) is beyond the GL paradigm, and the Landau order parameter
will acquire an enormous anomalous dimension. This phenomenon is confirmed by unbiased quantum Monte Carlo simulations~\cite{melko1,melko2}.
In the non-interacting limit, the quantum critical point between
two different topological insulators is usually described by a
gapless Dirac/Majorana fermion, but the role of strong interaction
at this transition has not been fully explored, although we
understand that in some particular cases interaction can gap out
this quantum critical point and lead to a continuous curve
connecting the two sides of the phase
diagram~\cite{Fidkowski1,Fidkowski2}. Quantum phase transitions
between bosonic SPT states were even less studied, and it was
pointed out that most generally two bosonic SPT states can be
separated by an intermediate phase~\cite{Tarun_PRB2013,lulee}.

With this in mind, it will be of great interest to investigate a
concrete example where in a strongly correlated fermionic SPT
setup there is a purely interaction-driven phase transition
between a topological insulator and a quantum disordered phase.
Such a bona fide interaction-driven topological phase transition
will have no mean-field (non-interacting) correspondence and
provide the precious example of a controlled study of the
interplay between non-trivial band topology and strong electron
interaction. And this is what we will focus on in this paper.

Here, we provide a concrete simple interacting fermion model that
is studied by large-scale unbiased QMC simulations. The results of
this investigation provide us with the following desired
phenomena: A bona fide interaction-driven quantum phase transition
between topological insulator and a strongly interacting Mott
insulator (a quantum disordered phase). We find that this quantum
critical point is fundamentally different from the TI-to-trivial
quantum phase transition in the free fermion limit, in the sense
that the fermions never close their gap at the transition, but
emergent collective bosonic degrees of freedom become critical.
Thus we can view this transition as a transition between a bosonic
SPT state and a trivial bosonic Mott insulator. And we demonstrate
that at one special point, this transition is described by a
$(2+1)d$ $O(4)$ nonlinear sigma model with {\it exact} $SO(4)$
symmetry, and a topological term at {\it exactly} $\Theta = \pi$.
Moreover, we also employ the strange correlator proposed by
Ref.~\cite{YZYou2014_SC} and tested in
Ref.~\cite{HQWu2015,Wierschem2014a,Wierschem2014b,Ringel2015} to
diagnose the topological nature of the interaction-driven quantum
phase transition between topological insulator and the strongly
interacting Mott insulator.




\section{Model and numerical method}
\label{sec:KMH_SC}
\subsection{AA-stacked bilayer KMH model with inter-layer AFM coupling}
\label{sec:BiKMHmodel} 
\begin{figure}[tp!]
\centering
\includegraphics[width=\columnwidth]{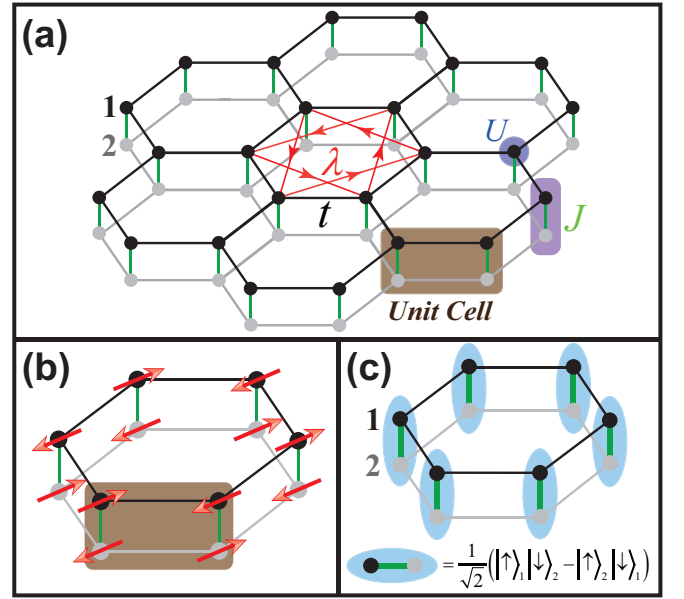}
\caption{\label{fig:HcLatt} (color online) (a) Illustration of
AA-stacked honeycomb lattice and bilayer KMH model with
inter-layer antiferromagnetic exchange interaction. The four-site
unit cell is presented as the shaded rectangle. The gray and black
lines indicates the nearest-neighbor hopping $t$ on layer 1 and 2,
respectively. The spin-orbital coupling term $\lambda$, for one
spin flavor, is shown by the red lines and arrows with
$\nu_{ij}=+1$. The on-site Coulomb repulsion and inter-layer AFM
coupling are represented by the shaded circle and rectangle, respectively. (b)
Illustration of the $xy$-AFM Mott insulator phase. (c)
Illustration of the inter-layer dimer-singlet phase. Shaded
ellipses are the inter-layer spin singlets.}
\end{figure}

In this work, we employ large-scale QMC
simulations to investigate the AA-stacked bilayer KMH model with
inter-layer AFM coupling, the Hamiltonian is given by, $\hat{H} = \hat{H}_{TB} + \hat{H}_U + \hat{H}_J$, as
\begin{eqnarray}
\hat{H} =&& -t\sum_{\xi\langle i,j \rangle,\alpha}( c^{\dagger}_{\xi i\alpha}c_{\xi j\alpha} + c^{\dagger}_{\xi j\alpha}c_{\xi i\alpha} )  \nonumber \\
&&+ i\lambda\sum_{\xi\langle\!\langle i,j \rangle\!\rangle, \alpha\beta}
v_{ij}(c^{\dagger}_{\xi i\alpha}\sigma^{z}_{\alpha\beta}c_{\xi j\beta} - c^{\dagger}_{\xi j\beta}\sigma^{z}_{\beta\alpha}c_{\xi i\alpha}) \nonumber \\
  &&+ \frac{U}{2}\sum_{\xi i}(n_{\xi i\uparrow} + n_{\xi i\downarrow} -1)^2 \nonumber \\
  &&+ \frac{J}{8}\sum_{i}\big[(D_{1i,2i}-D^{\dagger}_{1i,2i})^2 -(D_{1i,2i}+D^{\dagger}_{1i,2i})^2\big], \hspace{0.5cm}
\label{eq:ModelHamiltonian}
\end{eqnarray}
with $D_{1i,2i} =  \sum_{\sigma}c^{\dagger}_{1i\sigma}
c_{2i\sigma}$. $\alpha$, $\beta$ denote the spin species
$\uparrow$ and $\downarrow$ and $\xi=1,2$ stand for the layer
index in the AA-stacked bilayer system, as shown in
Fig.~\ref{fig:HcLatt}. $H_{TB}$ describes the tight-binding part
of the Hamiltonian, including the nearest-neighbor hopping and the
spin-orbit coupling~\cite{Kane2005a, Kane2005b} terms, and the
factor $v_{ij}=-v_{ji}=\pm1$ depends on the orientation of the two
nearest neighbor bonds that the electron traverses in going from
site $j$ to $i$, as shown in Fig.~\ref{fig:HcLatt} (a). The
$\sigma^{z}_{\alpha\beta}$ in the spin-orbit coupling term
furthermore distinguishes the $\uparrow$ and $\downarrow$ spin
states with the opposite next-nearest-neighbor hopping amplitude.
Throughout this work, we take $t$ as unit of energy. The second term
$H_{U}$ describes the on-site Coulomb repulsion between electrons,
and $n_{\xi i}=\sum_{\sigma}n_{\xi i\sigma}$. The electron filling
is fixed at half-filled, i.e., one electron per site on average.
The third term $H_{J}$ stands for the inter-layer
antiferromagnetic spin interaction. As explained in details in
Appendix~\ref{sec:appendix_a}, it is a faithful approximation of
the full Heisenberg interaction $J\sum_i\mathbf{S}_{1i}\cdot\mathbf{S}_{2i}$.

The Kane-Mele (KM) model preserves time-reversal symmetry $Z_2^T$
and its ground state is a quantum spin-Hall insulator with counter
propagating edge states~\cite{Kane2005a,Kane2005b}. On the
AA-stacked bilayer honeycomb lattice, the ground state of KM model
is still a QSH insulator but with two sets of counter-propagating
edge modes. As for the symmetry, the model Hamiltonian in
Eq.~(\ref{eq:ModelHamiltonian}) has charge $U(1)\times U(1)$
symmetry, which corresponds to charge conservation on each
individual layer. The $SU(2)$ spin-rotational symmetry is broken
down to $U(1)$ by the spin-orbit coupling term, the residual
$U(1)$ spin symmetry corresponds to the spin rotation in the $xy$
plane. Therefore, most generally the total symmetry of the
AA-stacked bilayer model is $U(1)_\text{spin}\times [U(1)\times
U(1)]_\text{charge} \rtimes Z_2^T$, which results in a
$\mathbb{Z}$ classification. This is because in the noninteracting limit we can define a Chern number for spin-up and spin-down electrons separately, and time-reversal symmetry guarantees that these two Chern numbers must be equal. Thus eventually the whole system is characterized by one Chern number, which can take arbitrary integer values. When including interaction terms, we found that at the limit of $U=0$, Eq.~(\ref{eq:ModelHamiltonian}) has a much higher $SO(4)$ symmetry, which we will analyze in detail in Sec.~\ref{sec:results_TPT}.

With interactions, the KMH model on the monolayer honeycomb
lattice has been studied by the Hartree-Fock mean-field
theory~\cite{Rachel2010}, cluster (dynamic) mean-field
theory~\cite{JianXinLi2011,Wu2012,Chen2015} as well as
determinantal QMC
simulations~\cite{Hohenadler2011,DZheng2011,Hohenadler2012,Assaad2013,Hung2013,Lang2013,Hung2014,
Meng2014,HQWu2015}. For the bilayer model in
Eq.~(\ref{eq:ModelHamiltonian}), at the $U=J=0$ limit, the system
is a QSH insulator with spin Chern number
$C_s=(C_{\uparrow}-C_{\downarrow})/2=2$ where $C_{\uparrow}=+2$
and $C_{\downarrow}=-2$ are the Chern numbers for spin-up and
spin-down parts. In presence of finite interactions, i.e., the
$U-J$ phase diagram, one can expect that in the large $U$ limit,
the bilayer system will be driven from the QSH state into a
$xy$-AFM ordered Mott insulator phase, through a continuous phase
transition, similar to the KMH model on monolayer honeycomb
lattice~\cite{Rachel2010,JianXinLi2011,Wu2012,Hohenadler2011,DZheng2011,Hohenadler2012,Assaad2013,Hung2013,
Lang2013,Hung2014,Meng2014,Chen2015,HQWu2015}, and the phase
transition should belong to the $(2+1)d$ XY universality
class~\cite{Hohenadler2012,Meng2014}. At large $J$ limit, the
bilayer system should enter the inter-layer dimer-singlet phase
with spin singlets formed on the inter-layer bonds due to strong
antiferromagnetic coupling $J$. In the $J\to\infty$ limit, the
inter-layer dimer-singlet phase is a product state of the
inter-layer singlets in which all the symmetries are
preserved~\cite{Slagle2015}. Combining the spin-orbit coupling
term $\lambda$, on-site Coulomb repulsion $U$ and inter-layer
coupling $J$, one can expect very interesting competition occurring among
the QSH, $xy$-AFM and inter-layer dimer-singlet phases, and it is the quantum phase transitions between these phases (some
of which is of exotic topological nature) that we engaged great
effort to unravel in this paper with unbiased large-scale QMC
simulations.

\subsection{Projector quantum Monte Carlo method}
\label{sec:PQMC}

Projector QMC (PQMC) method is the zero-temperature version of
determinantal QMC algorithm~\citep{AssaadEvertz2008}. PQMC method
obtains the ground-state expectation values of physical quantities
by carrying out an imaginary-time evolution of some trial
wavefunction, which is not orthogonal to the true many-body ground
state. The ground-state expectation value of physical observable
is calculated as follows,
\begin{eqnarray}
\langle \hat{O}\rangle =\lim\limits_{\Theta\to +\infty}
\frac{\langle
\psi_T|e^{-\Theta\hat{H}/2}\hat{O}e^{-\Theta\hat{H}/2}|\psi_T\rangle}{\langle
\psi_T|e^{-\Theta \hat{H}}|\psi_T\rangle},
\label{eq:PQMC_Observable}
\end{eqnarray}
where $|\psi_T\rangle$ is the trial wave function and $\Theta$ is
projection parameter. In all the simulations, to ensure that the
algorithm arrives at the truly converged ground state of finite
size systems, we choose $\Theta=60/t, \Delta\tau=0.05/t$, in which
$\Delta\tau$ is the finite imaginary-time step applied in the
Trotter decomposition of partition function. During the
simulations, we adopt the Hubbard-Stratonovich (HS) transformation
with four-component Ising fields to decouple the interaction
terms~\cite{Assaad2005}. Due to the fact that the two terms in
$H_J$ interaction do not commute, the systematic error for all
physical observables is at the order of $\mathcal{O}(\Delta\tau)$ (Trotter Error).
During the simulation, we make sure the value of $\Delta\tau$ is
small enough and the QMC sampling of physical observables is large
enough such that the results are numerically exact within
well-characterized statistical errors. We have simulated different
linear system size $L=3,6,9,12$ ($L=15$ for strange correlator),
with $N=L^2$ as number of unit cells to extrapolate physical
observables to the thermodynamic limit.

To determine the phase diagram for the bilayer model in
Eq.~(\ref{eq:ModelHamiltonian}), we first measure static physical
quantities, such as the expectation values of energy densities
(both total and each individual term in the Hamiltonian), double
occupancy, and spin-spin correlation function. The $xy$-plane AFM
order is expected to have ordering vector
$\boldsymbol{\Gamma}=(0,0)$~\cite{Hohenadler2011,DZheng2011,Hohenadler2012,Assaad2013,Hung2013,Lang2013,Hung2014,
Meng2014,Chen2015,Slagle2015,HQWu2015}, the transverse magnetic
structure factor at $\boldsymbol{\Gamma}$ point is measured as,
\begin{eqnarray}
S^{xy}(\boldsymbol{\Gamma})=\frac{1}{4N}\sum_{ij\gamma}\langle
S_{i\gamma}^xS_{j\gamma}^x+S_{i\gamma}^yS_{j\gamma}^y \rangle,
\label{eq:PQMC_StrucFac}
\end{eqnarray}
where $i,j=1,2,\cdots,N$ run over all unit cells and
$\gamma=1,2,3,4$ stands for the four sublattices inside a unit
cell. The staggered magnetic moment $m_S$ can be evaluated as
$m_S=\sqrt{S^{xy}(\boldsymbol{\Gamma})/N}$.

Next, to have the dynamical information of the system, such as the
excitation gaps in single- and two-particle channels, we need to
measure the imaginary-time single-particle Green's function,
\begin{equation}
G(\mathbf{k},\tau) =
\frac{1}{4N}\sum_{ij\gamma\sigma}e^{i\mathbf{k}\cdot
(\mathbf{R}_i-\mathbf{R}_j)}\langle
c^{\dagger}_{i\gamma\sigma}(\tau)c_{j\gamma\sigma} \rangle,
\label{eq:PQMC_GreenFucTau}
\end{equation}
where $\gamma$ is again the sublattice index and $\sigma$ is the
electron spin, and the imaginary-time spin-spin correlation
function at $\boldsymbol{\Gamma}$ point,
\begin{equation}
S^{xy}(\boldsymbol{\Gamma},\tau) =
\frac{1}{4N}\sum_{ij\gamma}\langle
S_{i\gamma}^x(\tau)S_{j\gamma}^x+S_{i\gamma}^y(\tau)S_{j\gamma}^y
\rangle,
\label{eq:PQMC_SpinPMTau}
\end{equation}
and the imaginary-time inter-layer pair-pair correlation
function in the charge channel
\begin{equation}
P(\boldsymbol{\Gamma},\tau) = \frac{1}{2N}\sum_{ij\delta}\langle \Delta^{\dagger}_{i\delta}(\tau)\Delta_{j\delta} + \Delta^{\dagger}_{j\delta}(\tau)\Delta_{i\delta}\rangle
\end{equation}
where $\Delta_{i\delta} =
\frac{1}{\sqrt{2}}(c_{1,i,\uparrow,\delta}c_{2,i,\downarrow,\delta}-c_{1,i,\downarrow,\delta}c_{2,i,\uparrow,\delta})$
is the inter-layer Cooper pair operator, it is defined on the two
inter-layer bonds $\delta=1,2$ of each unit cell $i$. At the
$\tau\to\infty$ limit, we access the asymptotic behavior
$G(\mathbf{k},\tau)\propto e^{-\Delta_{sp}(\mathbf{k})\tau}$,
$S^{xy}(\boldsymbol{\Gamma},\tau)\propto e^{-\Delta_{S}\tau}$ and
$P(\boldsymbol{\Gamma},\tau)\propto e^{-\Delta_{C}\tau}$ in which
$\Delta_{sp}(\mathbf{k})$ is the single-particle excitation gap and
$\Delta_{S}$, $\Delta_{C}$ are the two-particle excitation gaps
in the spin and charge channels for the interacting
system~\cite{Meng2010}. In our bilayer system, the minimum value
of single-particle gap appears either at $\mathbf{k}=\mathbf{K}$ or
$\mathbf{k}=\mathbf{M}$ depending on the parameters $U$ and $J$, and we
measure the spin and charge gaps at $\boldsymbol{\Gamma}$ point as it is the
ordered wave vector for the gapless Goldstone modes.

To diagnose the topological nature of the quantum phase transition,
we employ the recently developed strange correlator
method~\cite{YZYou2014_SC,HQWu2015,HQWu2015,Wierschem2014a,Wierschem2014b,Ringel2015}.
In the single-particle and two-particle (spin) channel, the correlation functions
constructed as,
\begin{equation}
C^{\sigma}_{\mathbf{k}AB}= \frac{\langle\Omega|c_{\mathbf{k}A\sigma}^{\dagger}c_{\mathbf{k}B\sigma}|\Psi\rangle}{\langle\Omega|\Psi\rangle}; \hspace{0.6cm}
S^{\pm}_{\mathbf{k}AA} = \frac{\langle\Omega|S^{+}_{\mathbf{k}A}S^{-}_{\mathbf{k}A}|\Psi\rangle}{\langle\Omega|\Psi\rangle},
\label{eq:StrCorrSP}
\end{equation}
where $c^{\dagger}_{\mathbf{k}A\sigma}=\frac{1}{L}\sum_{\xi,i}e^{i\mathbf{k}\cdot\mathbf{R}_{\xi i}}c^{\dagger}_{\xi i,A,\sigma}$ and $S^{+}_{\mathbf{k}A}=\frac{1}{L}\sum_{\xi,i}e^{i\mathbf{k}\cdot\mathbf{R}_{\xi i}}S^{+}_{\xi i,A}$ (integer $i$ as unit cell index),
with $\mathbf{k}$ inside the BZ region, $A$, $B$ are the sublattices in a
unit cell in one layer and $\xi$ the layer index, as shown in Fig.~\ref{fig:HcLatt}.
The basic idea of the strange correlator is
that, on the left hand side of the correlation function, the wave
function $|\Omega\rangle$ is a trivial band insulator (with spin
Chern number $C_{s}=0$); on the right hand side of the correlation
function, the projection operator $e^{-\Theta \hat{H}}$ guarantees
$|\Psi\rangle=e^{-\Theta \hat{H}}|\Psi_{T}\rangle$ is the
many-body ground state wave function of bilayer KMH Hamiltonian at
certain $J$ and $U$. If $|\Psi\rangle$ is topologically nontrivial QSH state, i.e., there exit
gapless edge modes at the spatial boundary of $|\Psi\rangle$, then after a space-time rotation,
$C^{\sigma}_{\mathbf{k}AB}$ will develop a singularity at certain symmetric momentum point $\mathbf{k}_s$: $C_{\mathbf{k}}$ $\sim$ $1/|\mathbf{k}-\mathbf{k}_s|^\alpha$, with $\alpha=1$ for
noninteracting system, $\alpha<1$ otherwise. Based on the
effective Lorentz invariant description of topological
insulators~\footnote{It is well-known that most topological
insulators can be described by Dirac fermions at low energy, and
the bosonic SPT states can be described by either a nonlinear
sigma model field theory~\cite{ZBi2015}, or a Chern-Simons field
theory~\cite{YMLu2012}, both of which have an effective Lorentz
invariance.}, the 2D strange-correlator $C^{\sigma}_{\mathbf{k}AB}$ should behave very similarly to the
$(1+1)d$ correlation functions at the boundary, endowed with a
Luttinger liquid description in the presence of interaction. If
$|\Psi\rangle$ is on the other hand a topological trivial
insulator, then the divergence in $C^{\sigma}_{\mathbf{k}AB}$ is
no longer present as there is no single-particle edge modes on the
boundary of $|\Psi\rangle$. What's more, the spin strange correlator $S^{\pm}_{\mathbf{k}AA}$ also has different behaviors, depending on whether the gapless two-particle edge modes is present or not. For QSH insulator, $S^{\pm}_{\mathbf{\Gamma}AA}$ should possess a diverging behavior faster than $\sim$ $\ln L$ (the case in noninteracting system) with increasing system size $L$, while it should saturate to finite value (slower than $\sim$ $\ln L$ behavior) in a topological trivial insulator. Thus, one can readily detect
the topological phase transition in the system by monitoring the
behavior of $C^{\sigma}_{\mathbf{k}AB}$ and $S^{\pm}_{\mathbf{\Gamma}AA}$. The strange correlation has been
successfully applied in the QMC investigation of the topological
phase transitions in the monolayer KMH model, the readers are
referred to Ref.~\cite{HQWu2015} for more details in its
physical meaning and technical implementation.


\section{Numerical results and discussions}
\label{sec:results}
\subsection{Phase diagram}
\label{sec:results_Phase} 
\begin{figure}[h]
\centering
\includegraphics[width=\columnwidth]{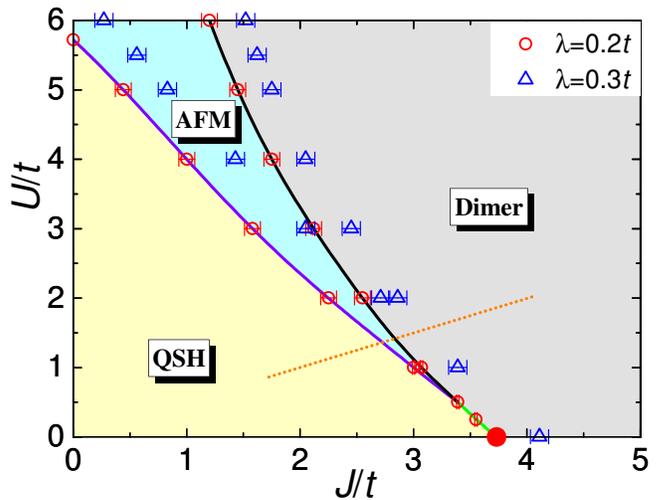}
\caption{\label{fig:Phase_Diagram}(color online) $U$-$J$ phase
diagram for the AA-stacked  bilayer KMH model with inter-layer
antiferromagnetic coupling. Showing here are the phase diagram for
$\lambda=0.2t$ and $\lambda=0.3t$ cases. Solid lines (violet, green
and black) are the phase boundaries for the $\lambda=0.2t$ case.
The red solid dot at $(J_c, U=0)$ and red open dots at $U=0.25$ and $0.5$ and the green line goes through them highlights the interaction-driven topological phase transition
between QSH and the dimer-singlet insulator phase. The orange dotted line highlights the $J=2U$ path which is studied in Ref.~\onlinecite{Slagle2015}, it actually goes through a small AFM region.}
\end{figure}

The $U-J$ phase diagram for
$\lambda=0.2t, 0.3t$ is shown in Fig.~\ref{fig:Phase_Diagram}, and this
is one of the main results of the paper. QSH, $xy$-AFM and
inter-layer dimer-singlet phases are found from QMC simulations.
Since there is only a net shift in the phase boundaries between
$\lambda=0.2t$ and $0.3t$ cases, we will focus on
the detailed results for the $\lambda=0.2t$ case in the following. The orange dotted line in Fig.~\ref{fig:Phase_Diagram} denotes the $J=2U$ path which is studied in Ref.~\cite{Slagle2015}, we note that with more careful finite size scaling in this work, we found it actually goes through an intermediate AFM region.

Three featuring observations about this phase diagram are in
order.

First of all, at small $U$ ($U < 0.5t$ for $\lambda=0.2t$) there
is a direct phase transition from the QSH insulator to inter-layer
dimer-singlet insulator (see details in Appendix~\ref{sec:appendix_d}). Notice that since neither the QSH nor the
dimer-singlet phase has symmetry breaking, all the symmetries
(such spin-rotation, charge conservation, time-reversal, and
lattice symmetry, etc.) in the model Hamiltonian
Eq.~(\ref{eq:ModelHamiltonian}) are preserved across this phase
transition, rendering it a bona fide topological phase transition
driven purely by the inter-layer antiferromagnetic interaction
$J$. This is a very unique case and very different from the
transitions in (interacting) topological insulators that have been
studied
before~\cite{Rachel2010,Hohenadler2011,DZheng2011,JianXinLi2011,Wu2012,Hohenadler2012,Assaad2013,
Hung2013,Lang2013,Hung2014,Meng2014,Laubach2014,Chen2015,HQWu2015},
where the transitions are either driven by hopping parameters at
free-fermion
level~\cite{Hung2013,Hung2014,Meng2014,Laubach2014,Chen2015}, or
after the transition the symmetry that protects the non-trivial
band topology has been destroyed by
interactions~\cite{Rachel2010,Hohenadler2011,DZheng2011,JianXinLi2011,
Wu2012,Hohenadler2012,Assaad2013,Hung2013,Lang2013,Hung2014,Meng2014,Laubach2014,
Chen2015,HQWu2015}. The nature of this exotic transition will be
further discussed in the next section.

Secondly, the region of $xy$-AFM phase is greatly extended by an
interesting collaboration between the on-site Coulomb repulsion
$U$ and the inter-layer AFM coupling $J$. At $J=0$, for
$\lambda=0.2t$, the QSH to $xy$-AFM phase transition occurs at
$U\approx5.6(2)t$~\cite{HQWu2015}, but as $J$ increases, the phase
boundary between QSH and $xy$-AFM moves towards smaller $U$, which
means $J$ and $U$ both prefer the AFM state, until $J$ dominates
over $U$, after which the dimer-singlet phase takes over. The same
phenomena is also observed for $\lambda=0.3t$ case.
\begin{figure}[htp!]
\centering
\includegraphics[width=\columnwidth]{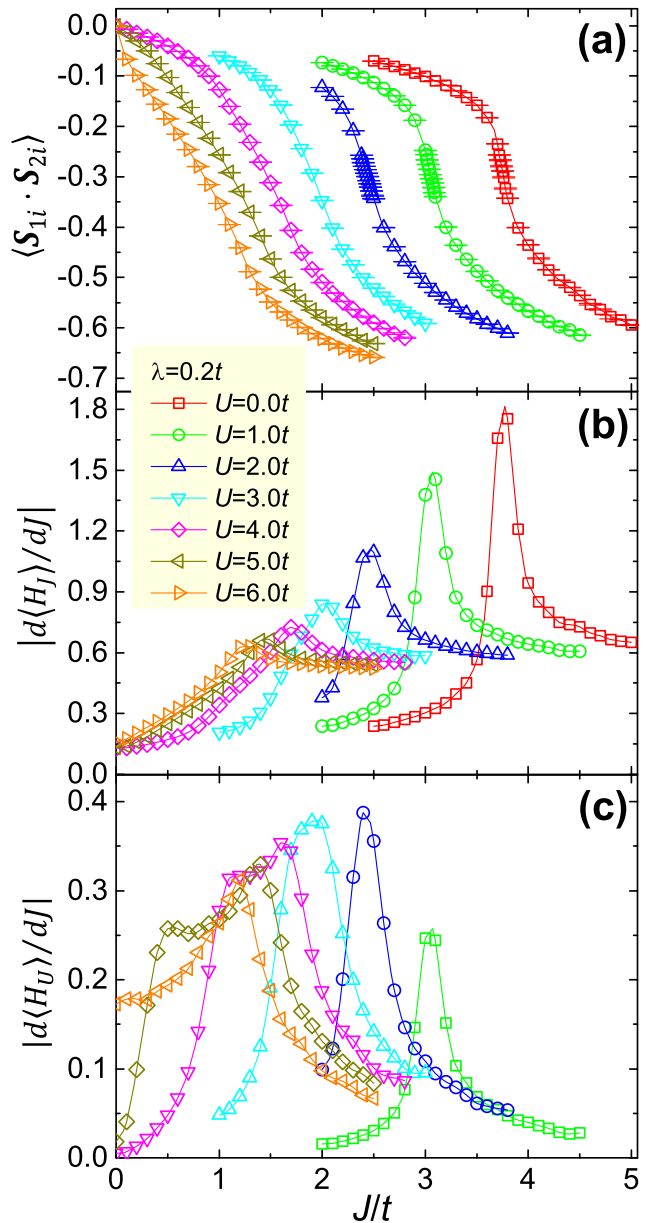}
\caption{\label{fig:HubbU_Diff}(color online) (a) The inter-layer
spin-spin correlation function for $L=6, \lambda=0.2t$ system with
various $U$ values, as a function of $J$. The continuous
variation of this correlation function indicates the topological
phase transition from QSH to dimer-singlet is a continuous one. At
large $J$, the correlation saturates at $-3/4$ which signifies the
formation of inter-layer dimer singlets. (b) First-order
derivative of $\langle H_J \rangle$ per bond over $J$ for $L=6,
\lambda=0.2t$. The peak in every curve explicitly indicates phase
transition from QSH insulator (or $xy$-AFM) phase to inter-layer
dimer-singlet phase. (c) First-order derivative of $\langle H_U
\rangle$ per site over $J$. The peaks in these curves indicate all
three possible phase transitions: QSH to dimer-singlet, QSH to
$xy$-AFM and $xy$-AFM to dimer-singlet transitions. For $U<2t$, the
peak in $d\langle H_U \rangle/dJ$ corresponds to the QSH to
dimer-singlet transition; for $U > 3t$, the two independent peaks
as a function of $J$ correspond to the QSH to $xy$-AFM transition
at small $J$ and $xy$-AFM to dimer-singlet transition at large
$J$.}
\end{figure}

Thirdly, for the direct phase transition from QSH phase to inter-layer
dimer-singlet phase, we have observed signatures of continuous phase transitions. This can be seen
from the inter-layer spin-spin correlation function per bond,
shown in Fig.~\ref{fig:HubbU_Diff} (a) for $L=6$ system with
$\lambda=0.2t$ (data with larger system sizes are shown in Appendix~\ref{sec:appendix_d}). For various $U$ values, as a function of $J$, the
spin-spin correlation function changes from 0 to $-3/4$, with the
latter signifying the formation of spin-singlet on every
inter-layer bond. Moreover, according to the Hellmann-Feynman
theorem, the spin-spin correlation function per bond is the
first-order derivative of the total energy density over $J$. Combining the results of $\langle\boldsymbol{S}_{1i}\cdot\boldsymbol{S}_{2i}\rangle$ presented in Fig.~\ref{fig:HubbU_Diff} (a) and in Appendix~\ref{sec:appendix_d}, the continuous changing of the first-order derivative of the total energy density, with increasing $J$, suggests that the topological phase transition from QSH to dimer-singlet insulator phase is continuous (at least for $U=0$).

To further elaborate upon this point, Fig.~\ref{fig:HubbU_Diff}
(b) and (c) show the first-order derivatives of expectation values
of $\langle H_J \rangle$ per bond and $\langle H_U \rangle$ per
site, over the parameter $J$. The peaks in
Fig.~\ref{fig:HubbU_Diff} (b) indicate the QSH to dimer-singlet
($U=0$) and $xy$-AFM to dimer-singlet (when $U\ge2t$) phase
transitions. The peaks in Fig.~\ref{fig:HubbU_Diff} (c) indicate
not only the same transitions in Fig.~\ref{fig:HubbU_Diff} (b) at
large $J$ and small $U$, but more interestingly, also the QSH to
$xy$-AFM phase transitions at small $J$ and large $U$ (for
$U\ge3t$), as there are two peaks in the curves for $U=3t,4t,5t$. The
finite-size effects in the energy density derivatives are small,
we only observe a slight shift of the phase boundaries for $L=9,
12$ systems, comparing with those for $L=6$ system shown here. In
the next section, we will present the finite-size scaling of the
QMC results of the magnetic order parameter as well as the
single-particle and spin excitation gaps across the topological phase
transition between QSH and dimer-singlet. As we will see, the
results hence obtained are consistent with those in
Fig.~\ref{fig:Phase_Diagram} and Fig.~\ref{fig:HubbU_Diff} in this
section.


\subsection{Topological phase transition}
\label{sec:results_TPT}

\subsubsection{Excitation gaps}
As mentioned in the preceding section
(Sec.~\ref{sec:results_Phase}), one of the most exciting features
in the phase diagram (Fig.~\ref{fig:Phase_Diagram}) is the exotic topological phase transition purely driven by the inter-layer
antiferromagnetic interaction $J$, between the QSH and
dimer-singlet phases.

\begin{figure}[tp!]
\centering
\includegraphics[width=\columnwidth]{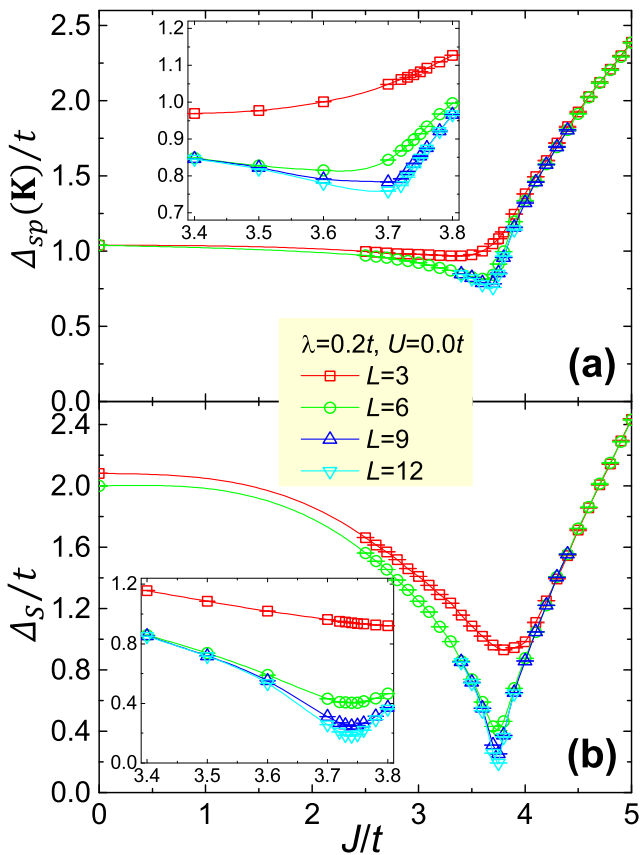}
\caption{\label{fig:Spin-SP_Gaps}(color online) (a)
Single-particle gap $\Delta_{sp}(\mathbf{K})$ of $\lambda=0.2t, U=0$ as a
function $J$. The inset shows the $\Delta_{sp}(\mathbf{K})$ in $J \in
[3.4t,3.8t]$ region. We have checked that $\mathbf{K}$ point is indeed the minimum of single-particle gap in the whole BZ. As a function of $J$, the single-particle gap only shows a gentle dip near the topological phase transition. (b) Spin gap $\Delta_S$ of
$\lambda=0.2t, U=0$ with increasing $J$. The inset is the spin gaps
in $J \in [3.4t,3.8t]$ region. The spin gap drops very fast and
closes at the topological phase transition point $J_c=3.73(1)t$.}
\end{figure}

In a free-fermion system, topological phase transitions between
SPT states are driven by tight-binding parameters. The
single-particle excitation gap will close to zero and reopen
continuously at the transition, as long as the symmetries
protecting the topologically nontrivial phase are still preserved.
However, the topological phase transitions in interacting systems
seems to be much more complicated. Of course, they can still be
driven by some tight-binding hopping parameter in the model
Hamiltonian, such as the third-nearest-neighbor
hopping~\cite{Hung2013, Hung2014,Meng2014,Chen2015}, dimerized
nearest-neighbor hopping~\cite{Lang2013,Meng2014,Hung2014}, Rashba
spin-orbit coupling~\cite{Laubach2014}, and Kekul\'e
distortion~\cite{Grandi2015} in the monolayer KMH model. In these
cases, single-particle gap closes and reopens at the topological
phase transition, just as their non-interacting counterparts. But,
they can furthermore be driven purely by interactions, such as
on-site Coulomb repulsion in monolayer KMH model~\cite{Rachel2010,
Hohenadler2011,DZheng2011,JianXinLi2011,Hohenadler2012,
Assaad2013,Meng2014,HQWu2015}, inter-layer AFM exchange coupling
in AA-stacked bilayer KMH model~\cite{Slagle2015}, and more
complicated form of interaction in interacting BHZ
model~\cite{Grandi2015}.

For interaction-driven topological phase transitions, the on-site
Coulomb repulsion in the monolayer KMH model drives the QSH phase
into an antiferromagnetically ordered phase with broken
time-reversal and spin rotational symmetries. Precisely speaking,
this is still not the topological phase transition we are after in
this paper: what we found here is an purely interaction-driven
topological phase transition without any symmetry breaking on
either side of the transition. Examples of this type of phase
transition has been discussed in 1-dimensional~\cite{Yoshida2014}
and 2-dimensional~\cite{Slagle2015} interacting systems.
In Ref.~\cite{Slagle2015}, the single-particle gap remains
gapped at the transition and it is the spin excitation gap and
Cooper pair gap that close and reopen. This implies that in the
low-energy limit such topological phase transition only involves
bosonic degrees of freedom, allowing the fermionic excitations to
be integrated out from the field theory~\cite{Slagle2015}.

Following Ref.~\cite{Slagle2015}, we perform a detailed
study on the topological phase transition between QSH and
dimer-singlet phases in the phase diagram of
Fig.~\ref{fig:Phase_Diagram}. To characterize this phase
transition, we measured the single-particle gap, two-particle spin
and charge gaps, as well as the strange
correlator~\cite{YZYou2014_SC,HQWu2015,Wierschem2014a,Wierschem2014b,Ringel2015}
in the QMC simulations.

\begin{figure}[t]
\centering
\includegraphics[width=\columnwidth]{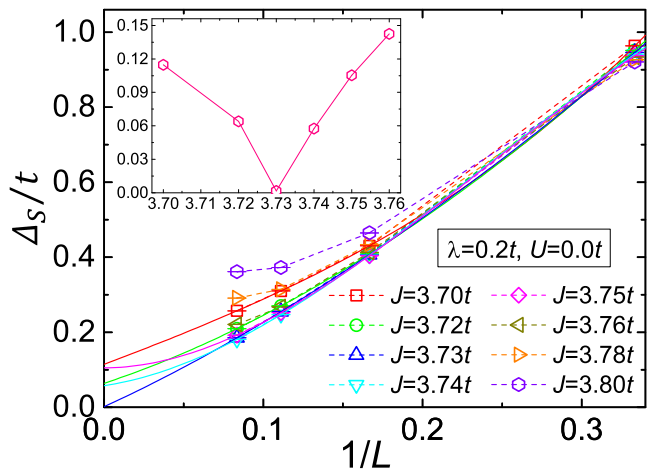}
\caption{\label{fig:Gap_Closing}(color online) Spin gap in
$J\in[3.7t,3.8t]$ region for $\lambda=0.2t, U=0$ with $L=3, 6, 9, 12$
and the extrapolation by third-order polynomial. The inset shows
the extrapolated spin gap as a function of $J$.}
\end{figure}

The results of single-particle and spin gaps with increasing $J$
are shown in Fig.~\ref{fig:Spin-SP_Gaps} for $\lambda=0.2t, U=0$ in
$L=3,6,9,12$ bilayer systems. The raw data of the single-particle
Green's function and dynamic spin-spin correlation function are
shown in Appendix~\ref{sec:appendix_b}, the data are of very good
quality, upon which we extracted the excitation gaps reliably. At
$U=0$, the topological phase transition point is at $J_c \simeq
3.7t-3.8t$ in the phase diagram in Fig.~\ref{fig:Phase_Diagram}. As
shown in Fig.~\ref{fig:Spin-SP_Gaps} (a), the single-particle gap
only exhibits a very gentle dip around the topological phase
transition point, which suggests the single-particle gap of the
system remains open as a function of $J$. In contrast, we observe
in Fig.~\ref{fig:Spin-SP_Gaps} (b) that the spin gap decreases
rapidly in the vicinity of $J_c$ as a function of system size $L$.
The inset of Fig.~\ref{fig:Spin-SP_Gaps} (b) shows the gap values
in the region $J\simeq3.4t-3.8t$. Within an even smaller region of
$J\in [3.7t,3.8t]$, we extrapolate the spin gap values for
$L=3,6,9,12$ systems in $1/L$ to estimate the spin gap in the
thermodynamic limit, which is shown in Fig.~\ref{fig:Gap_Closing}.
The main panel and inset of Fig.~\ref{fig:Gap_Closing} deliver a
clear message that the spin gap closing point is around
$J_c=3.73(1)t$. Furthermore, at $U=0$ as a function of $J$, we don't find a stepping of $xy$-AFM order by finite size extrapolation of the transverse magnetic structure factor (see details in Appendix~\ref{sec:appendix_d}).
At $J=3.8t$, the spin
gap values for $L=9$ system and $L=12$
are almost the same, indicating that the thermodynamic limit is
already reached and the spin excitations are well gapped here (spin-spin correlation in real space is exponentially short-ranged).
After the topological phase transition, the bilayer system enters
the inter-layer dimer insulator phase, which is schematically
shown in Fig.~\ref{fig:HcLatt} (c).

Combining the results for single-particle and spin gaps, we find
that the topological phase transition driven by inter-layer AFM
coupling in our bilayer system is fundamentally different from
those controlled by the hopping parameters, with and without
interactions~\cite{Rachel2010,
Hohenadler2011,DZheng2011,JianXinLi2011,Hohenadler2012,Wu2012,
Assaad2013,Hung2013,Lang2013,Meng2014,Hung2014,Meng2014,Laubach2014,Chen2015,HQWu2015}.

Also, at $U=0$, we have observed that the charge gap $\Delta_C$
and spin gap $\Delta_S$ are numerically identical (with difference
only up to $0.001t$). There is actually a deep theoretical reason
for the equality between these two-particle gaps: it is due to an
exact $SO(4)$ symmetry at $U=0$ (see Appendix
\ref{sec:appendix_c}), which rotates the $xy$-AFM (spin)
fluctuation $\vect{N}_i=\frac{1}{2}(-1)^{i+\xi-1}c_{\xi i}^\dagger\vect{\sigma}c_{\xi i}$ and the pairing (charge)
fluctuation $\Delta_i=c_{1i}\ii\sigma^y c_{2i}$ like an $O(4)$
vector:
\beqn
\vect{n}_i=(N^x_i, \Im\Delta_i, \Re\Delta_i, N^y_i).
\label{eq:O4vector}
\eeqn
Therefore both the spin and the charge
excitation gaps close identically at the transition point. To
better understand the $SO(4)$ symmetry, we may define two fermion
doublets $f_{i\sigma}$ ($\sigma = \uparrow, \downarrow$):
\beq
f_{i\uparrow}=\left(\begin{matrix}c_{1i\uparrow}\\(-1)^ic_{2i\uparrow}^\dagger\end{matrix}\right),
f_{i\downarrow}=\left(\begin{matrix}(-1)^i
c_{1i\downarrow}\\c_{2i\downarrow}^\dagger\end{matrix}\right).
\eeq
Then the $O(4)$ vector can be written as
\beq
\vect{n}_i = \frac{1}{2}f_{i\uparrow}^\dagger(\tau^0,-\ii\tau^1,-\ii\tau^2,-\ii\tau^3)f_{i\downarrow}+h.c.,
\eeq
where $\tau^{0,1,2,3}$ are the Pauli matrices acting on the
$f$-fermion doublets. The $SO(4)$ group is naturally factorized to
$SU(2)_\uparrow\times SU(2)_\downarrow$ as right and left
isoclinic rotations, under which the fermion transforms as
$f_{i\sigma}^{\dagger} \mapsto f_{i\sigma}^{\dagger}U_{\sigma}$ with $U_\sigma\in
SU(2)_\sigma$ for both $\sigma = \uparrow,\downarrow$. The model
Hamiltonian in \eqnref{eq:ModelHamiltonian} at $U=0$ can be
written in terms of the $SU(2)_\uparrow\times SU(2)_\downarrow$
singlets as
\begin{eqnarray}
\label{eq:BKMHQMCFfermion}
\hat{H} = \sum_{i,j,\sigma}\chi_{\sigma}(f_{i\sigma}^{\dagger}t_{ij}f_{j\sigma} + h.c.)
              - \frac{J}{4}\sum_{i}(\hat{P}^{\dagger}_{i}\hat{P}_{i}+ \hat{P}_{i}\hat{P}^{\dagger}_{i}), \hspace{0.6cm}
\end{eqnarray}
with
\begin{eqnarray}
\label{eq:POperator}
\hat{P}_i=\frac{1}{2}(-1)^{i}\sum_{\sigma}f_{i\sigma}^{\dagger}i\tau^2(f_{i\sigma}^{\dagger})^{\text{T}}
\end{eqnarray}
where we have $\chi_{\sigma}=(-1)^{\sigma}$, and $t_{ij}=t$ for hoppings on the NN bonds and $t_{ij}=i\lambda$ for SOC on the next-nearest-neighbor (NNN) bonds. Under arbitrary $SU(2)$ rotation of $f_{i\sigma}^{\dagger}$ operator as $f_{i\sigma}^{\dagger} \mapsto f_{i\sigma}^{\dagger}U_{\sigma}$, the $\hat{P}_i$ operator in Eq.~(\ref{eq:POperator}) is invariant since we have $f_{i\sigma}^{\dagger}i\tau^2(f_{i\sigma}^{\dagger})^{\text{T}} \mapsto f_{i\sigma}^{\dagger}U_{\sigma}i\tau^2U_{\sigma}^{\text{T}}(f_{i\sigma}^{\dagger})^{\text{T}}$, and the equality $U_{\sigma}i\tau^2U_{\sigma}^{\text{T}}=i\tau^2$ for $2\times2$ $SU(2)$ matrix $U_{\sigma}$. Besides, the hopping term $f_{i\sigma}^{\dagger}t_{ij}f_{j\sigma}$ is explicitly invariant under the $SU(2)_{\sigma}$ rotation of $U_{\sigma}$. Combining them, the Hamiltonian in Eq.~(\ref{eq:POperator}) has independent $SU(2)_{\uparrow}$ and $SU(2)_{\downarrow}$ symmetries for spin up and down channels, respectively. Thus, the $SO(4)\simeq SU(2)_{\uparrow}\times SU(2)_{\downarrow}$ symmetry for the bilayer model in Eq.~(\ref{eq:ModelHamiltonian}) under $U=0$ condition, which can be expressed in Eq.~(\ref{eq:BKMHQMCFfermion}), is explicit.

Physically, the $SO(4)$ symmetry rotates the four components of $\vect{n}_i$ defined in Eq.~(\ref{eq:O4vector}) to one another. As a result, the $xy$-AFM order should be exactly degenerate with the inter-layer spin-singlet $s$-wave superconducting order under $U=0$ condition due to the $SO(4)$ symmetry, which also indicates the identical excitation gaps corresponding to these two orders, i.e., spin gap and charge gap.

\subsubsection{Theoretical understanding}
In our phase diagram, the fermionic single-particle gap never
closes with finite $\lambda$, while the two-particle, collective,
bosonic modes (spin and charge gaps) both close at the
QSH-to-dimer-singlet phase transition, this means that at low
energy this model can be well-approximated by a bosonic model.
Indeed, Ref.~\cite{fb} demonstrated that many bosonic SPT
states can be constructed from fermionic topological
insulators/superconductors by confining the fermionic degrees of
freedom. In our case, we propose that the bosonic sector of our
phase diagram, at $U=0$, can be described by the following nonlinear sigma
model (NLSM) field theory~\cite{Slagle2015}: \beqn S =\int
\dd^2x\dd\tau \ \frac{1}{g} (\partial_\mu \vect{n})^2 +
\frac{\ii\Theta}{ \Omega_3}\epsilon_{abcd} n^a
\partial_x n^b
\partial_y n^c \partial_\tau n^d, \hspace{0.5cm}\label{o4nlsm}
\eeqn where $\Omega_3=2\pi^2$ is the volume of a three dimensional
sphere with unit radius. We will focus on the phase with large
$g$, namely the vector $\vect{n}$ is disordered.
Eq.~(\ref{o4nlsm}) is exactly the same field theory introduced by
Ref.~\cite{xusenthil,xuclass} to describe $2d$ bosonic SPT states,
and the physical meaning of the four component vector field
$\vect{n}$ was given in Eq.~(\ref{eq:O4vector}). As we show explicitly
in Appendix \ref{sec:appendix_c}, the model Eq.~(\ref{eq:ModelHamiltonian}) at $U=0$
has exactly $SO(4)$ symmetry, thus we do not need to turn on any
anisotropic term to Eq.~(\ref{o4nlsm}). When we move away from the
point $U=0$, an anisotropy needs to be turned on to split the
degeneracy between $(n_1,n_4)$ and $(n_2,n_3)$.

The phase diagram and renormalization group flow of the $(1+1)d$
analogue of \eqnref{o4nlsm} were calculated explicitly in
Refs.~\cite{pruisken1,pruisken2,pruisken2011}; and it was
demonstrated that the entire phase $0 \leq \Theta < \pi$ is
controlled by the trivial fixed point $\Theta = 0$, while the
entire phase $\pi < \Theta \leq 2\pi$ will flow to the fixed point
$\Theta = 2\pi$. The phase diagram of \eqnref{o4nlsm} was studied
in Ref.~\cite{xuludwig}, and again in the disordered phases
(phases with large $g$) $\Theta = \pi$ is the quantum phase
transition between the two phases with $0 \leq \Theta < \pi$ and
$\pi < \Theta \leq 2\pi$, the stable fixed point $\Theta = 2\pi$
describes a bosonic SPT state in $(2+1)d$~\cite{xuclass}.

The physical meaning of the fixed point $\Theta = 2\pi$ becomes
explicit when we create a vortex of $\Delta$, $i.e.$ the vortex of
$(n_2, n_3)$, then this vortex will acquire spin-1 due to the
$\Theta-$term at $\Theta=2\pi$, which is consistent with two
copies of quantum spin Hall insulator with $S^z$ conservation.
Also, at the fixed point $\Theta = 2\pi$, the boundary of
Eq.~(\ref{o4nlsm}) is a $(1+1)d$ $O(4)$ NLSM with a
Wess-Zumino-Witten term at level-1 \cite{xuludwig,xuclass}, whose
$SO(4)$ symmetry factorizes into $SU(2)_L \times SU(2)_R$ ($SU(2)$
symmetries for left and right moving modes respectively), where
$SU(2)_L$ and $SU(2)_R$ precisely correspond to $SU(2)_\uparrow$
and $SU(2)_\downarrow$ introduced in the previous subsection. Thus
the field theory Eq.~(\ref{o4nlsm}) does match with the all the
desired physics of our lattice model. In a later paper by some of us~\cite{You2016}, we demonstrate that the boundary state of our lattice model will be driven into a purely bosonic conformal field theory, in the sense that all the fermionic modes are gapped by interaction, but bosonic modes are gapless. And the remaining gapless bosonic modes at the boundary are precisely described by the boundary states of Eq.~(\ref{o4nlsm}). 

In Eq.~(\ref{o4nlsm}) $\Theta = \pi$ is the quantum phase transition
between the SPT and trivial phases, and in our phase diagram
$\Theta = \pi$ corresponds to the direct QSH-to-dimer-singlet
phase transition. Thus our lattice model actually provides a way to
simulate the topological field theory Eq.~(\ref{o4nlsm}) in QMC {\it
without sign-problem}.

\subsubsection{Strange correlator}
\label{sec:results_StrCorr} Let's now turn to understand the
topological phase transition from QSH to dimer-singlet phases from
the perspective of edge states. At $U=0$, in the QSH phase with
$J<J_c$, there exists two pair of gapless edge modes on the boundary of the bilayer KMH system, i.e., the spin
Chern number $C_s=2$. When $J>J_c$, the system is the
dimer-singlet state, it is a topologically trivial product state
hence the edge states are no longer present, i.e., spin Chern
number $C_s=0$. Therefore, the change of the topological nature
from QSH to dimer-singlet can be seen from the presence/absence of
the gapless edge states.

\begin{figure}[h!]
\centering
\includegraphics[width=\columnwidth]{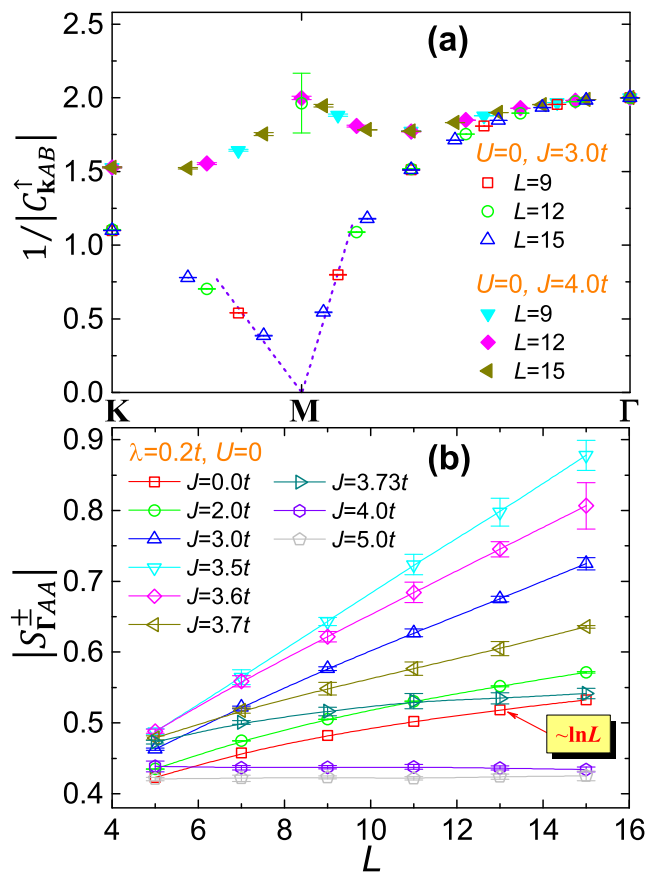}
\caption{\label{fig:strangecorrelator}(color online) (a) The inverse amplitude of single-particle strange correlator
$1/|C^{\uparrow}_{\mathbf{k}AB}|$ along the high-symmetry path for various $J=3.0t,4.0t$. (b) The spin strange correlator $S^{\pm}_{\mathbf{k}AA}$ at various $J$ values as a function of linear system size $L$.}
\end{figure}

In the QMC simulations, one can explicitly probe the spatial edge
by applying open boundary condition (OBC), but in interacting
systems, OBC usually has very strong finite-size dependence.
Moreover, to be able to see the edge mode, one further needs to
analytically continue the imaginary time correlation functions to have the spectra in real-frequency, but it is
well-known that analytical continuation usually generates
ambiguous results to the fine features of the spectra. Hence, to
avoid such difficulties, recently there is a new diagnosis dubbed
strange correlator, that has been proposed/tested successfully in
probing the edge states from static, bulk wave functions with
periodic boundary
condition~\cite{YZYou2014_SC,HQWu2015,Wierschem2014a,Wierschem2014b,Ringel2015}.

As explained in the Sec.~\ref{sec:PQMC}, whether the gapless edge modes is present in the bilayer system or not can be signified by the divergence of the
single-particle and spin strange correlator, which are shown in Fig.~\ref{fig:strangecorrelator} for $\lambda=0.2t, U=0$. From the single-particle strange correlator results in Fig.~\ref{fig:strangecorrelator} (a), for $J=3t$ ($J< J_c$), $|C^{\uparrow}_{\mathbf{k}AB}|$ of the bilayer
KHM mode is diverging at $\mathbf{M}$ point, correspondingly,
$1/|C^{\uparrow}_{\mathbf{k}AB}|$ vanishes in a power-law (the
exponent $\alpha$ is almost 1) to zero. The data point of
$1/|C^{\uparrow}_{\mathbf{k}AB}|$ exactly at $\mathbf{k}=\mathbf{M}$ is a
finite-size effect due to the implementation of strange correlator
in QMC and has been explained thoroughly in
Ref.~\cite{HQWu2015}. But when $J=4t$ ($J> J_c$), the divergence of
$C^{\uparrow}_{\mathbf{k}AB}$ is removed, hence the
$1/|C^{\uparrow}_{\mathbf{k}AB}|$ is no longer vanishing at $\mathbf{M}$
point, resembling the single-particle edge modes in QSH being gapped out due to
the inter-layer antiferromagnetic interaction $J$. As for the spin strange correlator shown in Fig.~\ref{fig:strangecorrelator} (b), $S^{\pm}_{\mathbf{k}AA}$ diverges with increasing $L$ at $J< J_c$, which is faster than the $\ln L$ behavior at the noninteracting limit $J=0$. These results indict the existence of gapless, spin (bosonic) edge modes~\cite{You2016}. On the contrary, $S^{\pm}_{\mathbf{k}AA}$ simply saturate to finite values when $J> J_c$, suggesting the absence of gapless edge modes. Combining the results of the strange correlator in both single-particle and two-particle channels, the QSH phase ($J< J_c$) in the bilayer model has gapless edge modes (bosonic), while they are absent in the dimer-singlet insulator phase, highlighting the topological phase transition.

\begin{figure}[b]
\centering
\includegraphics[width=\columnwidth]{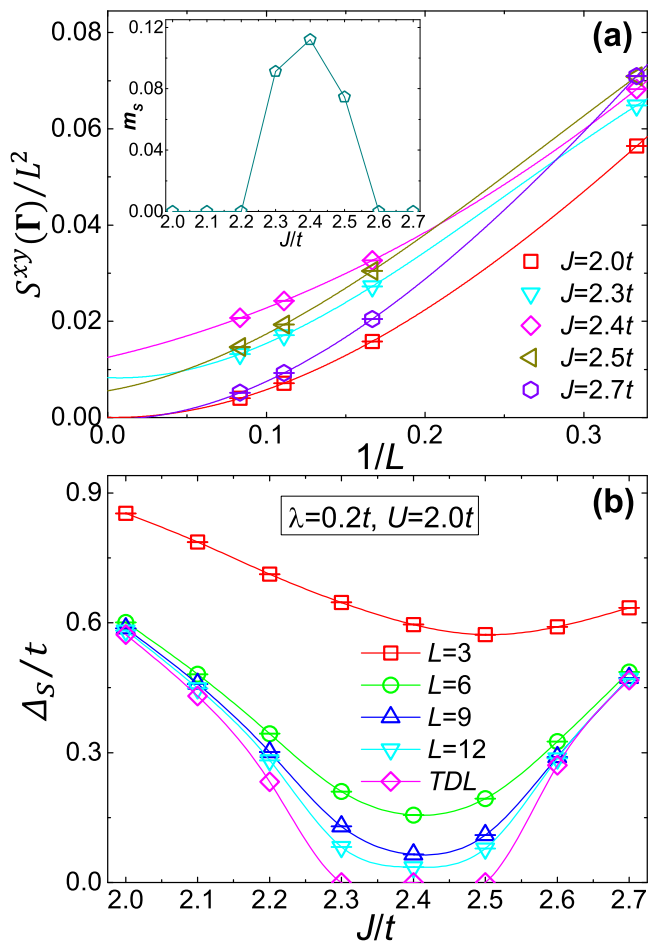}
\caption{\label{fig:U2_S_AF_Gap}(color online) (a) Finite-size
extrapolation of the transverse magnetic structure factor for
$L=3,6,9,12$ systems, the fits are third-order polynomial in
$1/L$. The parameters are $\lambda=0.2t, U=2t$ and $J \in
[2.0t,2.7t]$. Inset shows the extrapolated staggered magnetic moment
$m_{S}$ as a function of $J$. (b) The spin gap for $L=3,6,9,12$
systems and its extrapolated thermodynamic limit (TDL) values for
the same parameter set.}
\end{figure}

\subsection{$xy$-AFM order}
\label{sec:results_AFMOrder} The $xy$-AFM order in the phase
diagram Fig.~\ref{fig:Phase_Diagram} corresponds to the ordered
phase $g < g_c$ in Eq.~(\ref{o4nlsm}), with an extra anisotropy term
that favors $(n_1, n_2)$ over $(n_3, n_4)$. In the phase diagram
of Fig.~\ref{fig:Phase_Diagram}, one finds the region of $xy$-AFM
phase is greatly extended by an interesting collaboration between
the on-site Coulomb repulsion $U$ and the inter-layer AFM coupling
$J$. Intuitively, the $U$ term favors the $xy$-AFM state, while
the $J$ term favors the dimer-singlet state. With increasing $U$,
the QSH to $xy$-AFM and $xy$-AFM to dimer-singlet phase transition
points all move towards smaller $J$. This can be understood as
following: the $xy$-AFM phase is triggered by the intra-layer
antiferromagnetic coupling $J_\text{intra}\propto t^2/U$, the
dimer-singlet phase is triggered by the inter-layer $J$, their
phase transition is determined by the ratio $J/J_\text{intra}$,
since we get a smaller $J_\text{intra}$ for larger $U$, the
critical $J$ for the phase transition to dimer-singlet is
therefore reduced.

\begin{figure}[h!]
\centering
\includegraphics[width=\columnwidth]{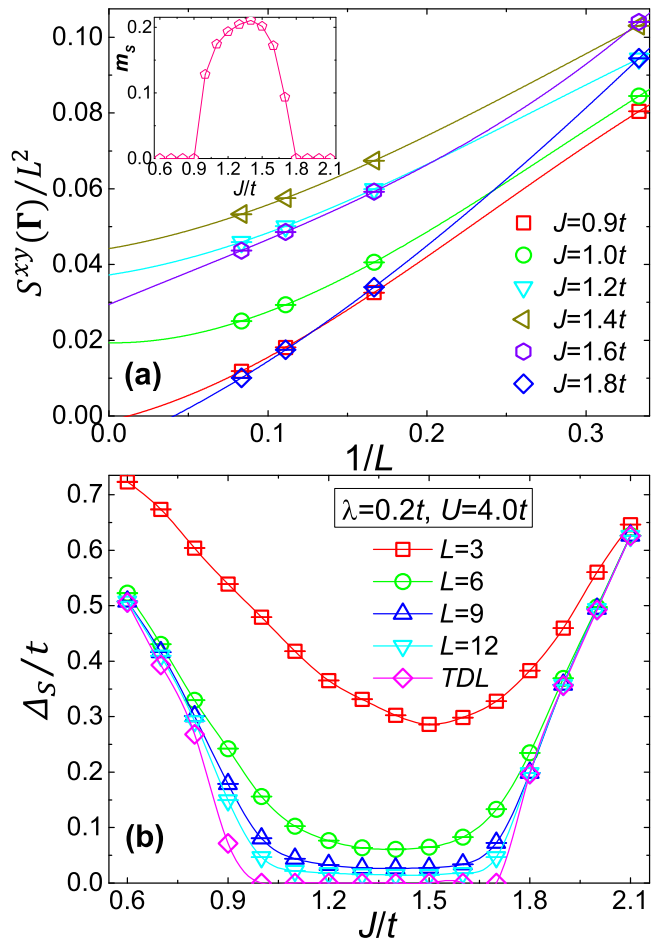}
\caption{\label{fig:U4_S_AF_Gap}(color online) (a) Finite-size
extrapolation of the transverse magnetic structure factor for
$L=3,6,9,12$ systems, the fits are third-order polynomial in
$1/L$. The parameter sets are $\lambda=0.2t, U=4t$ and
$J\in[0.6t,2.1t]$. Inset shows the extrapolated staggered magnetic
moment $m_{S}$ as a function of $J$. (b) The spin gap for
$L=3,6,9,12$ systems and its extrapolated thermodynamic limit
(TDL) values for the same parameter set.}
\end{figure}

Let us be more quantitative about the phase boundary. For the
monolayer KMH model with $\lambda=0.2t$, the system enters the
$xy$-AFM phase at $U_c=5.6(2)t$~\cite{HQWu2015}. In the presence of
inter-layer $J$, QMC results reveal that the $xy$-AFM phase can be
well established even at $U\sim2t$. As shown in
Fig.~\ref{fig:U2_S_AF_Gap} (a), for the magnetic structure factor
for $L=3,6,9,12$ systems and their extrapolation to thermodynamic
limit in $J\in [2.0t,2.7t]$, the extrapolated $S^{xy}(\boldsymbol{\Gamma})/N$ takes
nonzero values for $J=2.3t,2.4t,2.5t$ (see the inset of
Fig.~\ref{fig:U2_S_AF_Gap} (a)). To further confirm the long-range
magnetic order, we have also measured the spin gap and the results
are shown in Fig.~\ref{fig:U2_S_AF_Gap} (b). The extrapolated spin
gap at $J=2.3t,2.4t,2.5t$ are zero and corresponds to the Goldstone
mode associated with the $xy$-AFM long-range order. Combining the
data in Fig.~\ref{fig:U2_S_AF_Gap} (a) and (b), it's very
convincing that the long-range $xy$-plane magnetic order already
appears at $U\sim 2t$, almost 3 times smaller than that of the
$J=0$ case.

When the on-site Coulomb repulsion is further increased to $U=4t$,
at $\lambda=0.2t$ and $J \in [0.6t,2.1t]$, there are two phase
transitions (QSH to $xy$-AFM and $xy$-AFM to dimer-singlet) as $J$
increases. These can be detected by measuring the magnetic
structure factor and the spin gap as well, the results are show in
Fig.~\ref{fig:U4_S_AF_Gap}. Fig.~\ref{fig:U4_S_AF_Gap} (a) shows
that the system is in $xy$-AFM phase in $J\in[1.0t,1.7t]$ by finite
size extrapolation. The spin gap result in
Fig.~\ref{fig:U4_S_AF_Gap} (b) is well consistent with it, as the
spin excitations are gapless in the thermodynamic limit in
$J\in[1.0t,1.7t]$. When $J\le1.0t$, the system is inside the QSH
insulator where the spin excitations are gapped, and when
$J\ge1.7t$, the system is inside the dimer-singlet phase where the
spin excitations are gapped as well.

\section{SUMMARY AND OUTLOOK}

\label{sec:summary} In this work, we have found a bona fide
interaction-driven quantum phase transition between topological
insulator and a strongly interacting Mott insulator
(dimer-singlet). This quantum critical point is fundamentally
different from the TI-to-trivial quantum phase transition in the
non-interacting limit, in the sense that the fermions never close
their gap at the transition, instead, emergent collective bosonic
degrees of freedom become critical. We also employ the strange correlator proposed/tested in Ref.~\cite{YZYou2014_SC,HQWu2015,Wierschem2014a,Wierschem2014b,Ringel2015} to diagnose the topological nature of the quantum phase transitions.

In principle the exotic topological phase transition we found in
this paper can be generalized to all higher dimensions. What we
need to find is a higher dimension fermionic topological
insulator/superconductor that can be mapped to a bosonic SPT state
after confining the fermionic degrees of freedom, then in
principle the similar type of SPT-trivial phase transition with
gapless boson modes but no gapless fermion mode can be found in
these cases. A construction of these models in higher dimensions
was discussed in Ref.~\cite{fb}.

Although we have identified the field theory that describes this interaction-driven direct TI-to-trivial quantum phase transition in Eq.~(\ref{o4nlsm}), we do not yet have a controlled
analytical calculation for the universality class of this
transition. It seems the ordinary calculation techniques such
as $1/N$ or $\epsilon-$expansion both fail here, because
Eq.~(\ref{o4nlsm}) is defined solely for $(2+1)d$ and $O(4)$ vector.
How do we compare the critical scaling behavior of the spin gap $\Delta_S$ measured
in Fig.~\ref{fig:Gap_Closing} to theoretical calculations based on
Eq.~(\ref{o4nlsm}) is an interesting open question, which we will
leave to future study.


\begin{acknowledgments}

We acknowledge T. Yoshida, Z. Wang, K. Sun, N. Kawakami, X. Dai and F. Assaad for valuable discussions. The numerical calculations were carried out at the Physical Laboratory of High Performance
Computing in Renmin University of China, the Center
for Quantum Simulation Sciences in the Institute of Physics,
Chinese Academy of Sciences, the National Supercomputer
Center in Tianjin on the platform Tianhe-1A as well as the National Supercomputer Center in GuangZhou on the platform Tianhe-2A. YYH, HQW and ZYL acknowledge support from National Natural Science Foundation of China (NSFC Grant Nos. 11474356 and 11190024) and National Program for Basic Research of MOST of China (Grant No. 2011CBA00112). CX and YZY are supported by the David and Lucile Packard Foundation and NSF Grant No. DMR-1151208. ZYM is supported by the NSFC (Grant  Nos.  11421092  and  11574359) and the National Thousand-Young-Talents Program of China and acknowledges the hospitality of the KITP at the University of California, Santa Barbara, where part of this work is completed.

\end{acknowledgments}

\appendix
\section{approximate Heisenberg interaction}
\label{sec:appendix_a} In Sec.~\ref{sec:BiKMHmodel}, we mention
the inter-layer antiferromagnetic interaction in our Hamiltonian
is a faithful approximation of the full antiferromagnetic
Heisenberg interaction. Here we elaborate more upon this point.

The inter-layer interaction term $H_{J}$ in
Eq.~(\ref{eq:ModelHamiltonian}) can be written as summation of the
following term on all inter-layer bonds,
\begin{eqnarray}
\hat{Q}_i=\frac{1}{8}\left[(D_{1i,2i}-
D^{\dagger}_{1i,2i})^2-(D_{1i,2i}+D^{\dagger}_{1i,2i})^2\right].
\end{eqnarray}
There is an operator identity relates $\hat{Q}_i$ with full
Heisenberg exchange coupling~\cite{Assaad2005}, it reads
\begin{eqnarray}
\mathbf{S}_{1i}\cdot\mathbf{S}_{2i} = \hat{Q}_i -
\frac{1}{4}\Big[(\hat{n}_{1,i}-1)(\hat{n}_{2,i}-1)-1\Big],
\label{eq:SpinSpin_1}
\end{eqnarray}
so the difference between $\hat{Q}_i$ and
$\mathbf{S}_{1i}\cdot\mathbf{S}_{2i}$ is at the part
$[(\hat{n}_{1,i}-1)(\hat{n}_{2,i}-1)-1]$ (where indexes $1, 2$
stand for layers and integer $i$ for lattice site), but since our
system is half-filled, the expectation value of $\langle
\hat{n}_{1,i} \rangle = \langle \hat{n}_{2,i} \rangle = 1$, i.e.,
the charge fluctuations are small. This term can be safely
considered as a constant.



Moreover, it is easy to see $\mathbf{S}_{1i}\cdot\mathbf{S}_{2i}$ and
$\hat{Q}_i$ share the same eigenstates and their eigenvalues are
different only up to a $1/4$ shift. The eigenstates for
$\mathbf{S}_{1i}\cdot\mathbf{S}_{2i}$ and $\hat{Q}_i$ are spin
singlet and three-fold degenerate spin triplet states,
\begin{eqnarray}
|\psi_{0,+0}\rangle &=& \frac{1}{\sqrt{2}}\left(|\uparrow\rangle_1|\downarrow\rangle_2 - |\downarrow\rangle_1|\uparrow\rangle_2\right) \nonumber \\
|\psi_{1,+1}\rangle &=& \left(|\uparrow\rangle_1|\uparrow\rangle_2\right) \nonumber \\
|\psi_{1,+0}\rangle &=& \frac{1}{\sqrt{2}}\left(|\uparrow\rangle_1|\downarrow\rangle_2 + |\downarrow\rangle_1|\uparrow\rangle_2\right) \nonumber \\
|\psi_{1,-1}\rangle &=& \left(|\downarrow\rangle_1|\downarrow\rangle_2\right).
\end{eqnarray}
for $\mathbf{S}_{1i}\cdot\mathbf{S}_{2i}$, it's well known that
\begin{eqnarray}
\mathbf{S}_{1i}\cdot\mathbf{S}_{2i}|\psi_{0,0}\rangle &=& - \frac{3}{4}|\psi_{0,0}\rangle \nonumber \\
\mathbf{S}_{1i}\cdot\mathbf{S}_{2i}|\psi_{1,m}\rangle &=& + \frac{1}{4}|\psi_{1,m}\rangle\ \ \ \ m=0, \pm1.
\end{eqnarray}
for $\hat{Q}_i$ interaction, it's simple to show
\begin{eqnarray}
\hat{Q}_i|\psi_{0,0}\rangle &=& - 1\cdot|\psi_{0,0}\rangle \nonumber \\
\hat{Q}_i|\psi_{1,m}\rangle &=& + 0\cdot|\psi_{1,m}\rangle\ \ \ \ m=0, \pm1.
\end{eqnarray}





In terms of implementation in the PQMC simulations, for the $\hat{Q}_i$ term, we can directly apply
the following Hubbard-Stratonovich transformation to transform the
$\hat{Q}_i$ term into free fermion system coupled to 4-component
Ising fields,
\begin{eqnarray}
&&\exp\left[-\Delta\tau\frac{J}{8}(D_{1i,2i}-D^{\dagger}_{1i,2i})^2\right]  \nonumber \\
&=& \frac{1}{4}\sum_{l=\pm1,\pm2}\gamma(l)e^{i\xi_J\eta(l)(D_{1,2}-D^{\dagger}_{1,2})} + \mathcal{O}\left[(\Delta\tau)^4\right], \hspace{0.5cm} \nonumber \\
&&\exp\left[+\Delta\tau\frac{J}{8}(D_{1i,2i}+D^{\dagger}_{1i,2i})^2\right]  \nonumber \\
&=& \frac{1}{4}\sum_{l=\pm1,\pm2}\gamma(l)e^{\xi_J\eta(l)(D_{1,2}+D^{\dagger}_{1,2})} + \mathcal{O}\left[(\Delta\tau)^4\right], \hspace{0.5cm}
\end{eqnarray}
with $\xi_J=\sqrt{\Delta\tau J/8}$. For the full $\mathbf{S}_{1i}\cdot\mathbf{S}_{2i}$ interaction
term, we need to rewrite it into summation of $\hat{Q}_i$
interaction term, the on-site attractive interaction (the second term
in Eq.~\ref{eq:fullHeisenberg}) and the inter-layer
density-density attractive interaction (the third term Eq.~\ref{eq:fullHeisenberg}) as follows,
\begin{eqnarray}
\mathbf{S}_{1i}\cdot\mathbf{S}_{2i} = \hat{Q}_i &-& \frac{1}{4}\left[(\hat{n}_{1,i}-1)^2+(\hat{n}_{2,i}-1)^2\right] \nonumber \\
&-& \frac{1}{8}(\hat{n}_{1,i}+\hat{n}_{2,i}-2)^2.
\label{eq:fullHeisenberg}
\end{eqnarray}
The problem here is that with $J>0$ (the antiferromagnetic
interaction), the simultaneous presence of all these three terms will generate minus
sign problem to the QMC simulation under $U>0$ condition as the model in Eq.~(\ref{eq:ModelHamiltonian}), which effectively means that
there is no way to perform QMC simulation with the full Heisenberg interaction term for large systems. Though the QMC simulations applying the full $J$ term as Eq.~\ref{eq:fullHeisenberg} for the bilayer model under $U=0$ condition is free from sign problem, only keeping the $\hat{Q}_i$ in $\mathbf{S}_{1i}\cdot\mathbf{S}_{2i}$ interaction during the QMC simulations is still a good approximation, since the single-particle gap is always finite with $U=0$ and arbitrary $J$ parameter.




\section{Raw data for dynamic correlation functions}
\label{sec:appendix_b}

In Sec.~\ref{sec:results_TPT}, we present the single-particle as
well as the spin excitation gaps at the topological phase
transition between QSH and dimer-singlet phases. Here we show some
raw data for imaginary-time single-particle Green's function and
spin-spin correlation function, to provide the evidence that the
extrapolated excitation gaps are in good numerical quality.

\begin{figure}[h!]
\centering
\includegraphics[width=\columnwidth]{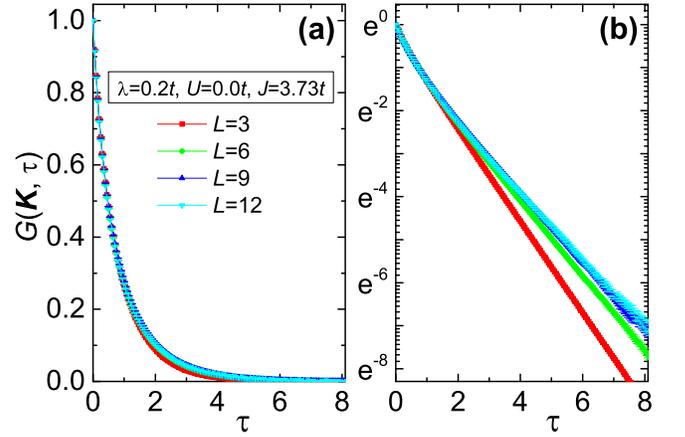}
\caption{\label{fig:U9_Single-particle-Gap_Data}(color online)
Single-particle Green's function for $\lambda=0.2t, U=0, J=3.73t$
with $L=3,6,9,12$ at $\mathbf{K}$ point in (a) linear scale and (b) in
semi-logarithmic scale.}
\end{figure}

\begin{figure}[h!]
\centering
\includegraphics[width=\columnwidth]{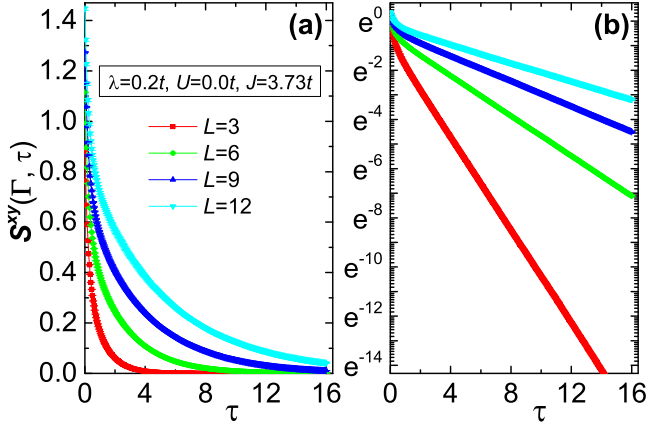}
\caption{\label{fig:U10_Spin-Gap_Data}(Color online) Dynamic
spin-spin correlation function for $\lambda=0.2t, U=0, J=3.73t$ with
$L=3, 6, 9, 12$ at $\boldsymbol{\Gamma}$ point in (a) linear scale and (b) in
semi-logarithmic scale.}
\end{figure}

Fig.~\ref{fig:U9_Single-particle-Gap_Data} and
Fig.~\ref{fig:U10_Spin-Gap_Data} are the raw data of the
single-particle Green's function $G(\mathbf{K},\tau)$ and the dynamic
spin-spin correlation function $S^{xy}(\boldsymbol{\Gamma},\tau)$, with
parameter set $\lambda=0.2t, U=0, J=3.73t$. According to
Fig.~\ref{fig:Spin-SP_Gaps}, this is exactly at $J=J_c$. In
Fig.~\ref{fig:U9_Single-particle-Gap_Data}(a), we can observe the
single-particle gap at $\mathbf{K}$ point decay very fast in imaginary time
$\tau$. In Fig.~\ref{fig:U9_Single-particle-Gap_Data} (b), with a
semi-logarithmic scale, we can see the size of the single-particle
gap almost converge to its thermodynamic limit value for $L=9, 12$
systems. Such fast decay and quick convergence with finite system
size actually means the single-particle gap is indeed finite and large at the topological phase transition. In fact it is about $0.7t$
at the transition point.

On the other hand, we can observe that the raw data for dynamic
spin-spin correlation function in Fig.~\ref{fig:U10_Spin-Gap_Data}
(a) decay slower with $\tau$. And in
Fig.~\ref{fig:U10_Spin-Gap_Data} (b) with a semi-logarithmic
scale, $S^{xy}(\boldsymbol{\Gamma},\tau)$ shows very good straight lines in
imaginary time $\tau$ and we can hence extract the spin gap value
with very high accuracy. In fact, the $1/L$ finite size scaling of
the spin gap at $J=3.73t$ gives rise a vanishing spin gap in the
thermodynamic limit.

\section{The $SO(4)$ symmetry}
\label{sec:appendix_c}
As mentioned in Sec.~\ref{sec:BiKMHmodel}, the  bilayer KMH model  given by
\eqnref{eq:ModelHamiltonian} has the
$U(1)_\text{spin}\times [U(1)\times U(1)]_\text{charge}\rtimes
Z_2^T$ symmetry in general. However when the model parameters are
tuned to certain special combinations, the model can have larger
symmetries. In this appendix, we will focus on the various unitary
symmetries of the model. The anti-unitary time-reversal symmetry
$Z_2^T$ is always presented and will be omitted in the following
discussion.

To understand the unitary symmetries systematically, let us first
introduce three sets of competing orders (in terms of fermion
bilinear operators): \beq
\begin{split}
\text{SDW: }&\vect{N}_i=(-1)^{\xi+i} c_{\xi i}^\dagger \vect{\sigma} c_{\xi i},\\
\text{SC: }&\Delta_i = c_{1 i}\,\ii\sigma^y c_{2 i},\\
\text{Exciton: }& D_i = (-1)^i c_{1 i}^\dagger c_{2 i},
\end{split}
\eeq where $c_{\xi i}=(c_{\xi i\uparrow},c_{\xi
i\downarrow})^\intercal$ is the fermion operator on site $i$ of
the $\xi$ layer. $(-1)^\xi$ and $(-1)^i$ respectively denote the
staggered sign factors between the layers and between the
sublattices. These competing orders anti-commute with each other,
and can be organized into an $O(7)$ vector:
$\vect{Q}_i=(N_i^x,N_i^y,N_i^z,\Re\Delta_i,\Im\Delta_i,\Re D_i,
\Im D_i)$. Then one can introduce the $SO(7)$ group on each site $i$
that rotates the vector $\vect{Q}_i$. The generators of the $SO(7)$
group are given by the following commutators (for $a<b$ and
$a,b=1,\cdots,7$) \beq \Gamma_i^{ab} =
\frac{1}{2\ii}[Q_i^a,Q_i^b]. \eeq The fermion operator transforms
under the $SO(7)$ rotation (parameterized by
$\theta_{ab}\in\mathbb{R}$) as \beq c_{\xi i \sigma} \to
\exp(\ii\theta_{ab}\Gamma^{ab}_i)c_{\xi i
\sigma}\exp(-\ii\theta_{ab}\Gamma^{ab}_i). \eeq The model
Hamiltonian in \eqnref{eq:ModelHamiltonian} can not achieve this
$SO(7)$ symmetry, but its achievable unitary symmetries are all
subgroups of this $SO(7)$. Different choices of the model parameters
breaks the $SO(7)$ symmetry differently.

To see how the $SO(7)$ symmetry is broken explicitly by the
Hamiltonian, we can calculate the commutator of the Hamiltonian
$H$ with the global $SO(7)$ generators
$\Gamma^{ab}\equiv\sum_i\Gamma_i^{ab}$: \beq
C^{ab}=\ii[H,\Gamma^{ab}].
\eeq
$C^{ab}=0$ means the Hamiltonian
has the symmetry that rotates $Q^a$ and $Q^b$. In general,
$C^{ab}$ is a linear combination of operators with the model
parameters $t$, $\lambda$, $U$ and $J$ as coefficients:
\beq
C^{ab}=t C^{ab}_t+\lambda C^{ab}_\lambda+U C^{ab}_U+J C^{ab}_J.
\eeq
$C^{ab}_t$, $C^{ab}_\lambda$, $C^{ab}_U$ and $C^{ab}_J$ are
complicated operators whose detail expressions are not of much
interest. We only need to extract the coefficients of linearly
independent operators, which are concluded in \tabref{tab:coefficients}.

Most generally, only 3 (out of 21) $SO(7)$ generators $\Gamma^{12}$,
$\Gamma^{45}$, $\Gamma^{67}$ commute with the Hamiltonian, as
$C^{12} = C^{45} = C^{67} = 0$. They generate the
$U(1)_\text{spin}\times[U(1)\times U(1)]_\text{charge}$
symmetry group. However, when $U=0$, we have
$C^{14}=C^{15}=C^{24}=C^{25}=0$ in addition, which enlarges the
symmetry group to $SO(4)\times U(1)$. The $SO(4)$ symmetry rotates
the $xy$-SDW order and the SC pairing order as an $O(4)$ vector $(N^x,
N^y, \Re\Delta,\Im \Delta)$, which involves particle-hole
transformations. The $U(1)$ symmetry rotates the exciton order $(\Re
D,\Im D)$ and corresponds to the conservation of the charge
difference between the layers. When $J=2U\neq 0$, we have
$C^{36}=C^{37}=0$, which enlarges the symmetry group to
SU(2)$\times$U(1)$_\text{spin}\times$U(1)$_\text{charge}$ as
mentioned in Ref.~\cite{Slagle2015}. When the interaction
is completely turned off as $U=J=0$, the model has $SO(4)\times SO(3)$
symmetry. On the other hand, in the absence of the spin-orbital
coupling, i.e. $\lambda = 0$, the model has even richer symmetry
structures, as the spin $SU(2)$ symmetry is restored. Under generic
interaction, the symmetry group is
$SU(2)_\text{spin}\times[U(1)\times U(1)]_\text{charge}$, which
can be enlarged to $SO(5)\times U(1)$ at $U=0$, or another
$SO(5)\times U(1)$ at $J=2U$, or $SO(4)\times SU(2)$ at $J=0$.

\begin{table}[h!]
\begin{center}
\begin{tabular}{cc|ccc|cc|cc|}
\cline{3-9}
&&\multicolumn{3}{c|}{SDW}&\multicolumn{2}{c|}{SC}&\multicolumn{2}{c|}{Exciton}\\
\cline{3-9}
& & $N^x$ & $N^y$ & $N^z$ & $\Re\Delta$ & $\Im \Delta$ & $\Re D$ & $\Im D$\\
\hline
\multicolumn{1}{|c|}{\multirow{3}{*}{SDW}} & $N^x$ & & $0$ & $\lambda$ & $U$ & $U$ & $J-2U$, $\lambda$ & $J-2U$, $\lambda$\\
\multicolumn{1}{|c|}{} & $N^y$ & & & $\lambda$ & $U$ & $U$ & $J-2U$, $\lambda$ & $J-2U$, $\lambda$\\
\multicolumn{1}{|c|}{} & $N^z$ & & & & $U$, $\lambda$ & $U$, $\lambda$ & $J-2U$ & $J-2U$\\
\hline
\multicolumn{1}{|c|}{\multirow{2}{*}{SC}} & $\Re\Delta$ & & & &  & $0$ & $J$, $\lambda$ & $J$, $\lambda$\\
\multicolumn{1}{|c|}{} & $\Im\Delta$ & & & &  & & $J$, $\lambda$ & $J$, $\lambda$\\
\hline
\multicolumn{1}{|c|}{\multirow{2}{*}{Exc.}} & $\Re D$ & & & &  &  & & $0$\\
\multicolumn{1}{|c|}{} & $\Im D$ & & & &  &  & & \\
\hline
\end{tabular}
\end{center}
\caption{Linearly independent coefficients in $C^{ab}$. For
example, in row $N^x$, column $\mathrm{Re}  D$, the number $J-2U$
and $\lambda$ mean that the commutation between $H$ and
$\frac{1}{2\ii}[N^x, \mathrm{Re} D]$ contains an operator with
coefficient $J-2U$, and another operator with coefficient
$\lambda$. The details of the form of the operators are not
shown.}
\label{tab:coefficients}
\end{table}

\section{The topological phase transitions at small $U$ region}
\label{sec:appendix_d}

\begin{figure}[tp!]
\centering
\includegraphics[width=\columnwidth]{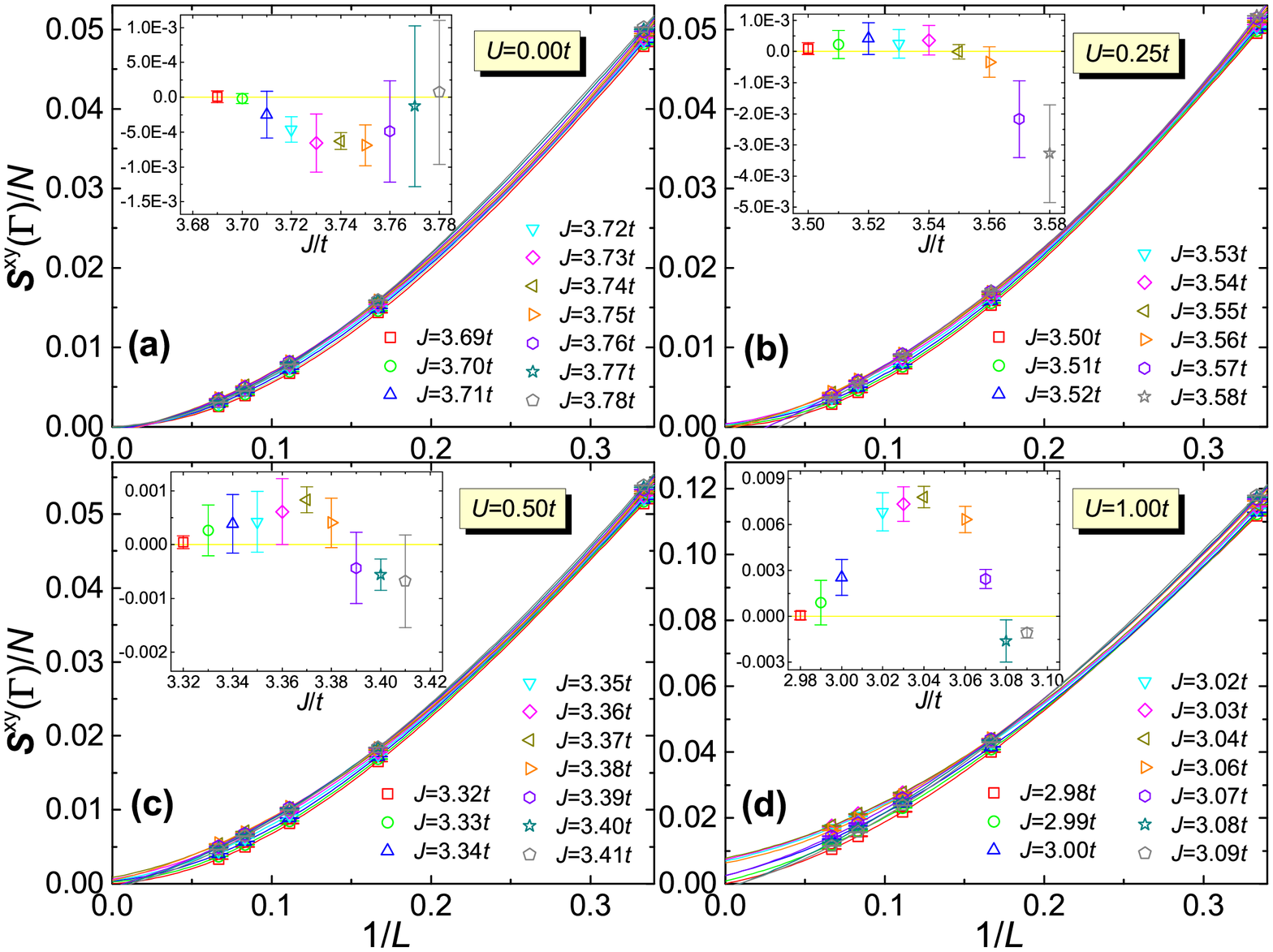}
\caption{\label{fig:SmallU_AFMOrder}(color online) Extrapolation of structure factor $S^{xy}(\boldsymbol{\Gamma})/N$ of $xy$-AFM order over $1/L$ for (a) $U=0$, (b) $U=0.25t$, (c) $U=0.50t$ and (d) $U=1.00t$ over inverse system size $1/L$, at $\lambda=0.2t$. The data points with error bars in the insets are the extrapolated values in the thermodynamic limit. }
\end{figure}

In Sec.~\ref{sec:results}, we present/discuss in detail the results about the $J$-driven topological phase transition without spontaneous symmetry breaking, including the energy derivatives, excitation gaps, strange correlator and quantum field theory correspondence. In this part, we show more numerical data about the topological phase transition at small $U$ region.

\begin{figure}[tp!]
\centering
\includegraphics[width=0.9\columnwidth]{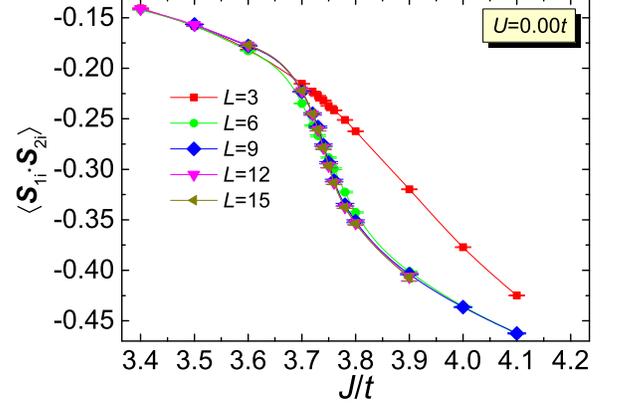}
\caption{\label{fig:EffectiveOrder}(color online) The inter-layer spin-spin correlation functions around the topological phase transitions for $U=0$ with $\lambda=0.2t$.}
\end{figure}

As we have mentioned, the $xy$-AFM order is absent around the topological phase transition at small $U$ region. In Fig.~\ref{fig:SmallU_AFMOrder}, the extrapolation of structure factors of $xy$-AFM order over $1/L$ for $U=0,0.25t,0.50t,1.0t$ are shown. From the results in Fig.~\ref{fig:SmallU_AFMOrder}, The $xy$-AFM order is explicitly absent for $U=0$ and $U=0.25$t, corresponding of which the topological phase transition points are $J_c/t=3.73t$ and $J_c/t=3.54t$. For $U=0.5t$, only a single point of $J/t=3.37$ has nonzero $xy$-AFM order applying the step size $\Delta J=0.01t$ during QMC simulations. Considering the numerical error existing in QMC simulations, it's reasonable to terminate the $xy$-AFM ordered phase at $U=0.5t$ in the phase diagram presented in Fig.~\ref{fig:Phase_Diagram}. For $U=1.0t$, the extrapolated $xy$-AFM order (inset of Fig.~\ref{fig:SmallU_AFMOrder} (d)) is nonzero in $3.00\le J/t\le 3.07$ region, which explicitly demonstrates the stepping in of $xy$-AFM ordered phase between the QSH insulator and inter-layer dimer-singlet insulator. Based on the results in Fig.~\ref{fig:SmallU_AFMOrder} (a), (b), the topological phase transition without spontaneous symmetry breaking is well established for $U=0$ and $U=0.25t$ case, i.e. finite $U$.

Another question is whether the topological phase transition at small $U$ is of first-order or continuous. Due to the fact that there is no nonzero local order parameter across the phase transition, to solve this problem thoroughly is not easy. However, to resolve this problem as best as we can, we have measured the inter-layer spin-spin correlation function $\langle\boldsymbol{S}_{1i}\cdot\boldsymbol{S}_{2i}\rangle$ for 5 different system sizes at $U=0$ as a function of $J$. As discussed in the main text, this quantity can be taken as the first-order derivative of ground state energy over $J$ parameter of the model in Eq.~(\ref{eq:ModelHamiltonian}). Depending on whether this quantity is continuous or not around the quantum phase transition in thermodynamic limit, we can determine the order of the transition.

The results of $\langle\boldsymbol{S}_{1i}\cdot\boldsymbol{S}_{2i}\rangle$ for $U=0$ across the topological phase transition are shown in Fig.~\ref{fig:EffectiveOrder}. We can observe that $\langle\boldsymbol{S}_{1i}\cdot\boldsymbol{S}_{2i}\rangle$ has almost reached the converged values already in $L=12$ system, i.e. values at thermodynamic limit. This is rather reasonable since the fermionic channel of the system is always gapped and the finite-size effect should should not be so strong. Most importantly, we indeed observe that $\langle\boldsymbol{S}_{1i}\cdot\boldsymbol{S}_{2i}\rangle$ changes smoothly across the topological phase transition, which suggests a continuous phase transition.

\bibliography{BilayerBib}

\end{document}